\documentclass[aps,prb,twocolumn,amsmath,amssymb,floatfix,superscriptaddress]{revtex4-1}

\usepackage{amsmath}
\usepackage{amsfonts}
\usepackage{amssymb}
\usepackage{graphicx}			
\usepackage{subfig}				
\usepackage{epstopdf}			
\usepackage{bm}				
\usepackage{units}				
\usepackage{dcolumn}			
\usepackage{verbatim}			
\usepackage[font=footnotesize]{caption}
\usepackage{xcolor}				
\begin{document}

\title{Generation and Evolution of High-Mach Number, Laser-Driven Magnetized Collisionless Shocks in the Laboratory}

\author{D. B. Schaeffer}
	\affiliation{Department of Astrophysical Sciences,  Princeton University, Princeton, NJ 08540, USA}
\author{W. Fox}
   	\affiliation{Princeton Plasma Physics Laboratory, Princeton, New Jersey 08543, USA}
\author{D. Haberberger}
   	\affiliation{Laboratory for Laser Energetics, University of Rochester, Rochester, New York 14623, USA}
\author{G. Fiksel}
   	\affiliation{Center for Ultrafast Optical Science, University of Michigan, Ann Arbor, MI 48109, USA}
\author{A. Bhattacharjee}
   	\affiliation{Department of Astrophysical Sciences,  Princeton University, Princeton, NJ 08540, USA}
	\affiliation{Princeton Plasma Physics Laboratory, Princeton, New Jersey 08543, USA}
\author{D. H. Barnak}
  	\affiliation{Laboratory for Laser Energetics, University of Rochester, Rochester, New York 14623, USA}
	\affiliation{Fusion Science Center for Extreme States of Matter, University of Rochester, Rochester, New York 14623, USA}
\author{S. X. Hu}
   	\affiliation{Laboratory for Laser Energetics, University of Rochester, Rochester, New York 14623, USA}
\author{K. Germaschewski}
   	\affiliation{Space Science Center, University of New Hampshire, Durham, New Hampshire 03824, USA}	

\maketitle


\textbf{
Shocks act to convert incoming supersonic flows to heat, and in collisionless plasmas the shock layer forms on kinetic plasma scales through collective electromagnetic effects.  These collisionless shocks have been observed in many space and astrophysical systems \cite{smith_jupiters_1975, smith_saturns_1980, burlaga_magnetic_2008, sulaiman_quasiperpendicular_2015}, and are believed to accelerate particles, including cosmic rays, to extremely high energies \cite{kazanas_origin_1986, loeb_cosmic_2000, bamba_small-scale_2003, masters_electron_2013, ackermann_detection_2013}.  Of particular importance are the class of high-Mach number, supercritical shocks \cite{balogh_physics_2013} ($M_A\gtrsim4$), which must reflect significant numbers of particles back into the upstream to accommodate entropy production, and in doing so seed proposed particle acceleration mechanisms \cite{blandford_particle_1978, mcclements_surfatron_2001, caprioli_simulations_2014-1, matsumoto_stochastic_2015}.  Here we present the first laboratory generation of high-Mach number magnetized collisionless shocks created through the interaction of an expanding laser-driven plasma with a magnetized ambient plasma.  Time-resolved, two-dimensional imaging of plasma density and magnetic fields shows the formation and evolution of a supercritical shock propagating at magnetosonic Mach number $M_{ms}\approx12$.  Particle-in-cell simulations constrained by experimental data show in detail the shock formation, separate reflection dynamics of C$^{+6}$ and H$^{+1}$ ions in the multi-species ambient plasma, and density and magnetic field compressions and overshoots in the shock layer.  The development of this experimental platform complements present remote sensing and spacecraft observations, and opens the way for controlled laboratory investigations of high-Mach number collisionless shocks, including the mechanisms and efficiency of particle acceleration.
}

\begin{figure}
	\centering
	\includegraphics{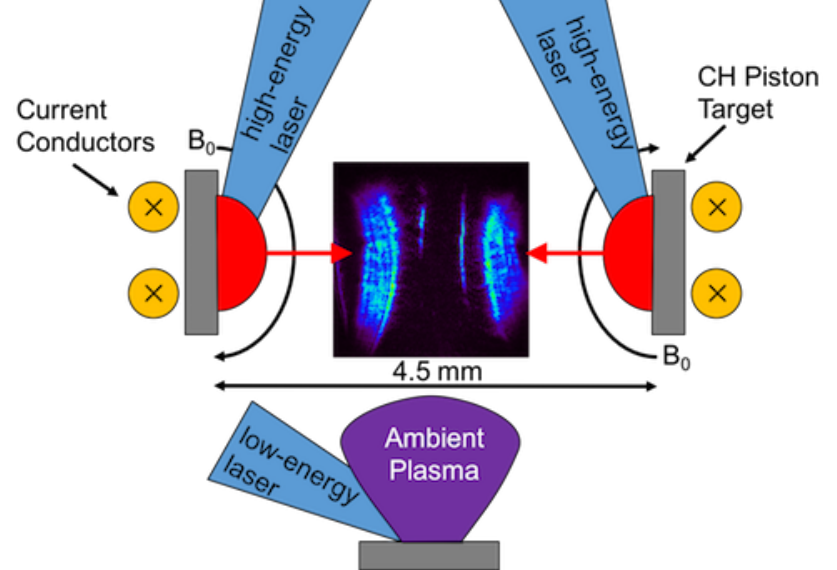}
	\caption{\textbf{Experimental setup.}  An external magnetic field ($B_0=8$ T) in an anti-parallel geometry was applied by pulsing current through conductors located behind opposing plastic (CH) piston targets.  The volume between the piston targets was pre-filled with an ambient plasma irradiated by a low-energy beam from a third CH target.  Two counter-propagating piston plasma plumes, generated by ablating the opposing piston targets with two high-energy beams, were then driven through the magnetized ambient plasma.  An example image taken with the angular filter refractometry diagnostic is shown for reference (see Methods).}
	\label{fig:setup}
\end{figure}

\begin{figure*}
	\centering
	\includegraphics{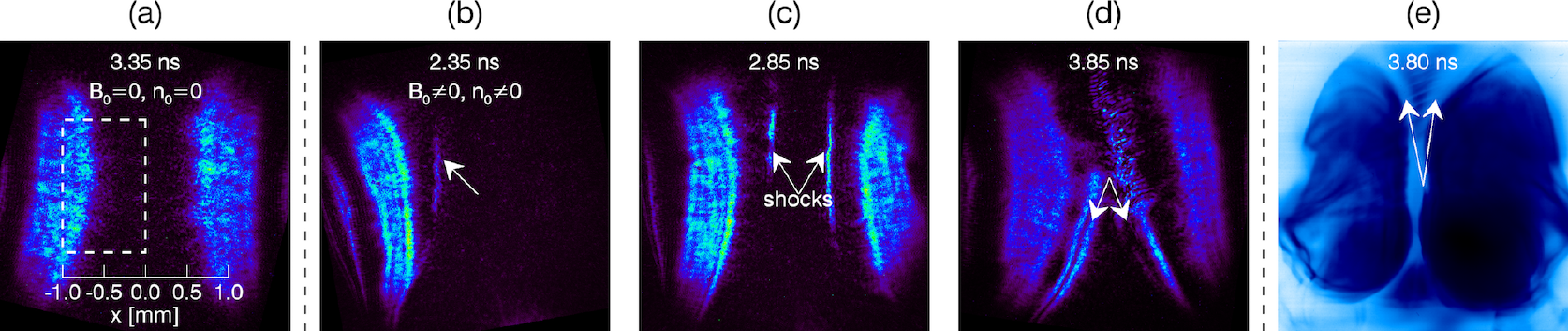}
	\caption{\textbf{Refractive and proton radiographic images of collisionless shock evolution.} In each image, the piston targets are located just outside the left and right borders, while the ambient target is located below the bottom border.  The piston plasmas expand toward the center ($x=0$), while the ambient plasma expands from the bottom upwards. The timestamps correspond to the time relative to the firing of the drive beams.  The dashed rectangle in (a) represents the region of interest in Fig. \ref{fig:oep_profiles}.  Panels (a)-(d) are images of angular filter refractometry.  In (a), no shock is observed without an external magnetic field or ambient plasma.  In (b)-(d), the shock is observed evolving from early to late times with an external field and ambient plasma.  In (b), only one target was used, confirming that these features are independent of counter-streaming interactions between two pistons.  In (e), proton radiography reveals the formation of strong magnetic field compressions (light ``ribbons'') coincident with the shock at comparable times.}
	\label{fig:oep_summary}
\end{figure*}

\begin{figure}
	\centering
	\includegraphics{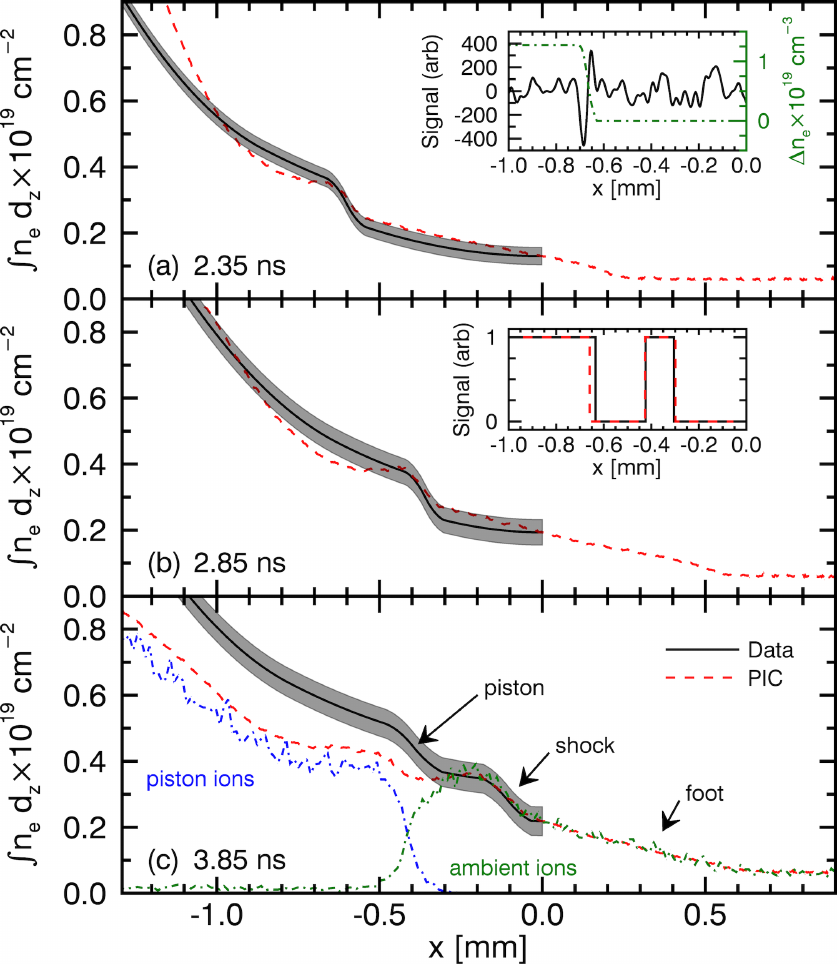}
	\caption{\textbf{Evolution of plasma density profiles.}   Line-integrated electron density profiles (black) were reconstructed from refractive imaging at (a) 2.35 ns, (b) 2.85 ns, and (c) 3.85 ns after laser ablation (see Methods).  For each, the shaded band corresponds to the uncertainty in the integration length.  Also shown are the corresponding profiles from \textsc{psc} PIC simulations (red).  Additionally, in (c) the ambient (green) and piston (blue) contributions to the total electron density in the PIC simulations are shown.  (a, inset) Raw shadowgraphy signal (black) and reconstructed relative density (green) profile at 2.35 ns. (b, inset) Direct comparison of the raw AFR signal (black) and corresponding synthetic simulation signal (red) at 2.85 ns.  For both, the signals have been reduced to binary for simplicity.  In all plots, the plasma moves toward $x=0$.}  
	\label{fig:oep_profiles}
\end{figure}

\begin{figure*}
	\centering
	\includegraphics{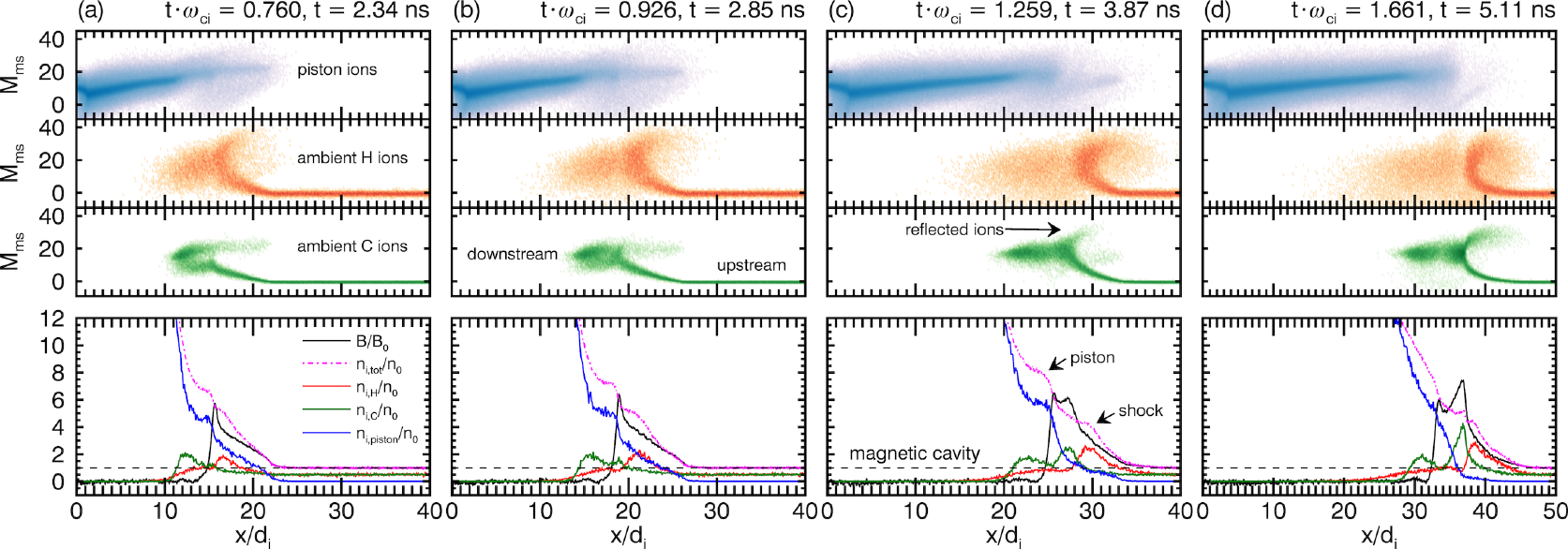}
	\caption{\textbf{Results from 2D particle-in-cell simulations that show the formation of a high-Mach number, magnetized collisionless shock.}  The simulations consist of a CH piston plasma expanding into a CH ambient plasma in an externally-applied magnetic field.  Each panel is representative of a different time in units of the upstream C ion gyroperiod $\omega_{ci}^{-1}$, form earliest (a) to latest (d).  For each panel, the top three rows are phase space density plots of piston, ambient H, and ambient C ions in terms of the C magnetosonic Mach number $M_{ms}$ and ambient C ion inertial length $d_i$.  The bottom row are plots of magnetic field (black), total ion density (purple), ambient H ion density (red), ambient C ion density (green), and piston ion density (blue) relative to their upstream ambient values.}
	\label{fig:psc_summary}
\end{figure*}

On October 7, 1962 the Mariner II spacecraft confirmed for the first time the existence of \textit{collisionless} shocks when it passed through the magnetic discontinuity that is  known today as the Earth's bow shock \cite{sonett_the-distant_1963}.  Such shocks had only been proposed a decade earlier \cite{de-hoffmann_magneto-hydrodynamic_1950}, and differ from hydrodynamic shocks in that they dissipate energy through collective electromagnetic effects on length scales far shorter than the classical mean free path.  Since then, collisionless shocks have been observed throughout the cosmos, including around Earth and planets of the solar system \cite{smith_jupiters_1975, smith_saturns_1980, masters_electron_2013, sulaiman_quasiperpendicular_2015}, the heliopause \cite{burlaga_magnetic_2008}, and supernovae remnants \cite{spicer_model_1990}.  In many of these systems, the shocks are \textit{magnetized} due to pre-existing magnetic fields in the upstream plasma.  In the heliosphere magnetized shocks have been well-studied within the limitations implied by 1-D spacecraft trajectories.  Meanwhile, remote sensing of supernovae provide compelling evidence of particle heating \cite{bamba_small-scale_2003} and cosmic-ray acceleration \cite{ackermann_detection_2013} typically attributed to shocks, but such observations are too distant to resolve details of the shock itself.  Consequently, laboratory experiments -- with their reproducibility and comprehensive, multi-dimensional datasets -- can complement spacecraft and remote sensing observations through an appropriate scaling of key dimensionless parameters \cite{drake_design_2000}.


The present experiments were carried out on the Omega EP laser facility \cite{waxer_high-energy_2005} at the University of Rochester.  The experimental setup is shown in Fig. \ref{fig:setup}.  Two opposing plastic (CH) piston targets were embedded in an externally applied magnetic field and irradiated by one or two high-energy lasers, driving counter-propagating, supersonic plasma plumes through a pre-formed, magnetized ambient CH plasma in a perpendicular magnetic geometry.  This experimental configuration utilizes the concept of a magnetic piston pioneered by early experiments \cite{paul_experimental_1965}, and has been previously used to demonstrate how expanding, laser-driven piston plumes sweep up and compress the ambient plasma and magnetic field \cite{fiksel_magnetic_2014}.  It also extends previous, low-Mach number shock experiments \cite{niemann_observation_2014, schaeffer_generation_2012, schaeffer_experimental_2015} to a significantly new parameter space, with magnetic fields, ambient densities, and laser energies that are orders of magnitude larger.  Other versions of this configuration have focused on the formation of non-magnetized (electrostatic) collisionless shocks \cite{romagnani_observation_2008, kuramitsu_time_2011, haberberger_collisionless_2011} or counter-streaming geometries relevant to Weibel-mediated shocks \cite{fox_filamentation_2013, ross_collisionless_2013}.

The interaction of the piston plasmas with the magnetized ambient plasma was diagnosed with a 263 nm probe beam that passed through the plasmas, producing simultaneous images of shadowgraphy and angular filter refractometry (AFR).  Additionally, the dynamics and topology of the magnetic fields were probed using a multi-MeV proton beam generated with an independent short-pulse laser (see Methods).


Fig. \ref{fig:oep_summary} shows AFR and proton radiographic images.  The main features seen in the AFR images include two wide bands, and, in some conditions, one or two very narrow bands near the image center.  These narrow bands are also seen in the same locations in the corresponding shadowgraphy images (not shown).  The wide bands are associated with the piston plasma plumes, while the narrow bands indicate the development of very strong density gradients where the piston and ambient plasmas interact.  In particular, Figs. \ref{fig:oep_summary}b-d show the formation and evolution of shock-like features from early to late times.  In contrast, Fig. \ref{fig:oep_summary}a shows that without an external magnetic field and without an ambient plasma, no shock forms (see Methods for additional null results).  We note that shock formation only requires a single piston plume interacting with the ambient plasma (see Fig. \ref{fig:oep_summary}b); multiple pistons were used to increase data collection.  Fig. \ref{fig:oep_summary}e is a proton radiographic image and shows the formation of strong magnetic field compressions (light regions of low proton fluence) coincident with the AFR shock bands, as well as the formation of magnetic cavities (dark regions) behind the magnetic compressions (see Methods).

The experiments were simulated with the 2D particle-in-cell (PIC) \textsc{psc} and 2D radiation-hydrodynamic \textsc{draco} codes (see Methods).  \textsc{draco} was used to model the laser-target interaction in order to predict the density and temperature profiles of the ambient and piston plasmas; the associated wide plume bands imaged through AFR were found to be in good agreement with these \textsc{draco} predictions (see Methods).  Modeling by \textsc{draco} further indicated that the ambient plasma electron density and temperature remain relatively stable at $n_{e,a0}\approx(0.2-0.6)\times10^{19}$ cm$^{-3}$ and $T_{e,a0}\approx15$ eV between the piston targets over the timescales of the experiment.  We used these plume parameters to initialize a fully kinetic \textsc{psc} simulation of the expansion of a mixed-species CH piston plasma into a uniform, pre-formed ambient CH plasma embedded in a uniform magnetic field.  In this configuration, the piston plasma drives a strong diamagnetic current and accompanying magnetic compression as it expands through the ambient plasma, coupling energy and momentum to the ambient ions through induced Larmor electric fields \cite{hewett_physics_2011}.  The magnetic pulse is then carried by the newly accelerated ambient ions and steepens into a collisionless shock through ion reflection.  

Density profiles reconstructed from the refractive images are shown in Fig. \ref{fig:oep_profiles}.  These profiles and associated numerical modeling show conclusive evidence of the production of a supercritical magnetized collisionless shock, as indicated by a magnetosonic Mach number $M_{ms}>4$, a density compression $n/n_0>2$, a compression ramp width $\Delta x/d_i>1$, and the separation of the shock structure from the piston.  The speed of the narrow bands in Fig. \ref{fig:oep_summary} can be estimated from their time-of-flight.  Between 2.35 and 2.85 ns (Fig. \ref{fig:oep_profiles}a-b), it is found to be $v_s=700\pm30$ km/s.  This implies that the bands are moving highly super-magnetosonically with $M_{ms}=v_s/c_{ms}=12\pm 5$, where $c_{ms}^2=v_A^2+c_s^2$, $v_A$ is the Alfv\'{e}n speed, $c_s$ is the sound speed, and both are calculated relative to conditions in the upstream (ahead of the shock) ambient C plasma.  Note that at these speeds, the ions are effectively collisionless (despite high densities), with a ratio of the collisional length scale $\lambda_{ii}$ to the system size $D_0$ (the distance between piston targets) of $\lambda_{ii}/D_0>1$.  As the bands form, their corresponding jump in density grows from $\Delta n_e=(1.3\pm 0.3)\times10^{19}$ cm$^{-3}$ at 2.35 ns to $\Delta n_e=(1.6\pm 0.3)\times10^{19}$ cm$^{-3}$ at 2.85 ns.  These density jumps represent a growth in the maximum compression ratio relative to the background of $n/n_0\geq3.2\pm0.5$ to $n/n_0\geq3.7\pm0.5$.  Simultaneously, the width of the density jump grows from $\Delta x=84\pm 10$ $\mu$m to $\Delta x=140\pm 10$ $\mu$m.  In terms of the ambient C ion inertial length $d_i$, this represents a growth from $\Delta x/d_i\geq0.6\pm0.2$ to $\Delta x/d_i\geq1.1\pm0.3$.  Here, the inequalities indicate that these measurements are lower bounds since the diagnostics are only sensitive to the largest density gradients, rather than the entire density jump.  Similar to previous experiments \cite{fiksel_magnetic_2014}, we conclude that the magnetic field is compressed by $B/B_0\gtrsim3$ (see Methods).

These features are confirmed by PIC simulations (see Fig. \ref{fig:psc_summary}a-b), which show that by 2.35 ns a collisionless shock has already formed in the ambient H ions and is moving at $M_{ms}\sim15$.  This can be seen by the large density and magnetic field compressions ($n/n_0>3$, $B/B_0>3$), the large compression ramp widths ($\Delta x/d_i>2$), a population of hot downstream H ions (in the shock frame, they are heated as they are decelerated through the shock front), and a small population of highly-accelerated ambient H ions (in the shock frame, these are reflected ions).  By 2.85 ns, the H shock has evolved to a stronger state (large compressions and reflected ion populations), while an additional shock is just forming in the ambient C ions.  Note that 2.35 ns and 2.85 ns correspond to H and C gyration times of $t\cdot\omega_{ci,H}\sim1.4$ and $t\cdot\omega_{ci,C}\sim0.9$, respectively, which is consistent with the expected shock formation timescale $t\cdot\omega_{ci}\sim1$ for each species.

By 3.85 ns (Fig. \ref{fig:oep_profiles}c), the data-derived density profile has bifurcated into double ``bumps''.  From shadowgraphy, the first bump is consistent with the one seen at 2.85 ns.  Simulations (Fig. \ref{fig:psc_summary}c) reveal that by this time the C shock has fully formed, while the H shock has maintained its structure in a quasi-steady state.  The formation of the C shock has also, in particular, forced the piston ions to begin piling-up as they become trapped behind the C-generated magnetic compression.  As a result, a double bump structure forms in the density profile due to the separation of the C (and H) shock from the pile-up of piston ions behind it.  At later times (Fig. \ref{fig:psc_summary}d), the C shock becomes the dominant feature, with significantly stronger downstream and reflected ion populations compared to the H shock.  By these late times the piston ions also become effectively trapped behind the C shock, indicating that their role in coupling energy and momentum to the ambient ions has effectively ended.


These results demonstrate that we can drive high-Mach number, magnetized collisionless shocks in the laboratory, and open a new experimental regime for studying shock formation and evolution that is difficult to achieve with spacecraft.  The experimental platform is highly flexible -- allowing variation in the applied magnetic field, upstream density, magnetic geometry, and piston speed -- and its development enables new collaborative investigations on the relationship between collisionless shocks and other highly-driven, astrophysically-relevant systems such as those involving magnetic reconnection or the Weibel instability.


\bibliographystyle{naturemag}
\bibliography{omega-cs_biblib}

\begin{thebibliography}{10}
\expandafter\ifx\csname url\endcsname\relax
  \def\url#1{\texttt{#1}}\fi
\expandafter\ifx\csname urlprefix\endcsname\relax\def\urlprefix{URL }\fi
\providecommand{\bibinfo}[2]{#2}
\providecommand{\eprint}[2][]{\url{#2}}

\bibitem{smith_jupiters_1975}
\bibinfo{author}{Smith, E.~J.} \emph{et~al.}
\newblock \bibinfo{title}{Jupiter's magnetic field. magnetosphere, and
  interaction with the solar wind: Pioneer 11}.
\newblock \emph{\bibinfo{journal}{Science}} \textbf{\bibinfo{volume}{188}},
  \bibinfo{pages}{451--455} (\bibinfo{year}{1975}).

\bibitem{smith_saturns_1980}
\bibinfo{author}{Smith, E.~J.} \emph{et~al.}
\newblock \bibinfo{title}{Saturn's magnetic field and magnetosphere}.
\newblock \emph{\bibinfo{journal}{Science}} \textbf{\bibinfo{volume}{207}},
  \bibinfo{pages}{407--410} (\bibinfo{year}{1980}).

\bibitem{burlaga_magnetic_2008}
\bibinfo{author}{Burlaga, L.~F.} \emph{et~al.}
\newblock \bibinfo{title}{Magnetic fields at the solar wind termination shock}.
\newblock \emph{\bibinfo{journal}{Nature}} \textbf{\bibinfo{volume}{454}},
  \bibinfo{pages}{75--77} (\bibinfo{year}{2008}).

\bibitem{sulaiman_quasiperpendicular_2015}
\bibinfo{author}{Sulaiman, A.~H.} \emph{et~al.}
\newblock \bibinfo{title}{Quasiperpendicular high mach number shocks}.
\newblock \emph{\bibinfo{journal}{Phys. Rev. Lett.}}
  \textbf{\bibinfo{volume}{115}}, \bibinfo{pages}{125001}
  (\bibinfo{year}{2015}).

\bibitem{kazanas_origin_1986}
\bibinfo{author}{Kazanas, D.} \& \bibinfo{author}{Ellison, D.~C.}
\newblock \bibinfo{title}{Origin of ultra-high-energy $\gamma$-rays from cygnus
  {X}-3 and related sources}.
\newblock \emph{\bibinfo{journal}{Nature}} \textbf{\bibinfo{volume}{319}},
  \bibinfo{pages}{380--382} (\bibinfo{year}{1986}).

\bibitem{loeb_cosmic_2000}
\bibinfo{author}{Loeb, A.} \& \bibinfo{author}{Waxman, E.}
\newblock \bibinfo{title}{Cosmic $\gamma$-ray background from structure
  formation in the intergalactic medium}.
\newblock \emph{\bibinfo{journal}{Nature}} \textbf{\bibinfo{volume}{405}},
  \bibinfo{pages}{156--158} (\bibinfo{year}{2000}).

\bibitem{bamba_small-scale_2003}
\bibinfo{author}{Bamba, A.}, \bibinfo{author}{Yamazaki, R.},
  \bibinfo{author}{Ueno, M.} \& \bibinfo{author}{Koyama, K.}
\newblock \bibinfo{title}{{Small-Scale} strcuture of the 1006 shock with
  chandra observations}.
\newblock \emph{\bibinfo{journal}{The Astrophysical Journal}}
  \textbf{\bibinfo{volume}{589}}, \bibinfo{pages}{827--837}
  (\bibinfo{year}{2003}).

\bibitem{masters_electron_2013}
\bibinfo{author}{Masters, A.} \emph{et~al.}
\newblock \bibinfo{title}{Electron acceleration to relativistic energies at a
  strong quasi-parallel shock wave}.
\newblock \emph{\bibinfo{journal}{Nat Phys}} \textbf{\bibinfo{volume}{9}},
  \bibinfo{pages}{164--167} (\bibinfo{year}{2013}).

\bibitem{ackermann_detection_2013}
\bibinfo{author}{Ackermann, M.} \emph{et~al.}
\newblock \bibinfo{title}{Detection of the characteristic pion-decay signature
  in supernova remnants}.
\newblock \emph{\bibinfo{journal}{Science}} \textbf{\bibinfo{volume}{339}},
  \bibinfo{pages}{807--811} (\bibinfo{year}{2013}).

\bibitem{balogh_physics_2013}
\bibinfo{author}{Balogh, A.} \& \bibinfo{author}{Treumann, R.~A.}
\newblock \emph{\bibinfo{title}{Physics of Collisionless Shocks}},
  vol.~\bibinfo{volume}{12} of \emph{\bibinfo{series}{ISSI Scientific Report
  Series}} (\bibinfo{publisher}{Springer}, \bibinfo{year}{2013}).

\bibitem{blandford_particle_1978}
\bibinfo{author}{{Blandford}, R.~D.} \& \bibinfo{author}{{Ostriker}, J.~P.}
\newblock \bibinfo{title}{{Particle acceleration by astrophysical shocks}}.
\newblock \emph{\bibinfo{journal}{The Astrophysical Journal}}
  \textbf{\bibinfo{volume}{221}}, \bibinfo{pages}{L29--L32}
  (\bibinfo{year}{1978}).

\bibitem{mcclements_surfatron_2001}
\bibinfo{author}{McClements, K.~G.}, \bibinfo{author}{Dieckmann, M.~E.},
  \bibinfo{author}{Ynnerman, A.}, \bibinfo{author}{Chapman, S.~C.} \&
  \bibinfo{author}{Dendy, R.~O.}
\newblock \bibinfo{title}{Surfatron and stochastic acceleration of electrons at
  supernova remnant shocks}.
\newblock \emph{\bibinfo{journal}{Phys. Rev. Lett.}}
  \textbf{\bibinfo{volume}{87}}, \bibinfo{pages}{255002}
  (\bibinfo{year}{2001}).

\bibitem{caprioli_simulations_2014-1}
\bibinfo{author}{Caprioli, D.} \& \bibinfo{author}{Spitkovsky, A.}
\newblock \bibinfo{title}{Simulations of ion acceleration at non-relativistic
  shocks. i. acceleration efficiency}.
\newblock \emph{\bibinfo{journal}{The Astrophysical Journal}}
  \textbf{\bibinfo{volume}{783}}, \bibinfo{pages}{91} (\bibinfo{year}{2014}).

\bibitem{matsumoto_stochastic_2015}
\bibinfo{author}{Matsumoto, Y.}, \bibinfo{author}{Amano, T.},
  \bibinfo{author}{Kato, T.~N.} \& \bibinfo{author}{Hoshino, M.}
\newblock \bibinfo{title}{Stochastic electron acceleration during spontaneous
  turbulent reconnection in a strong shock wave}.
\newblock \emph{\bibinfo{journal}{Science}} \textbf{\bibinfo{volume}{347}},
  \bibinfo{pages}{974--978} (\bibinfo{year}{2015}).

\bibitem{sonett_the-distant_1963}
\bibinfo{author}{Sonett, C.~P.} \& \bibinfo{author}{Abrams, I.~J.}
\newblock \bibinfo{title}{The distant geomagnetic field: 3. disorder and shocks
  in the magnetopause}.
\newblock \emph{\bibinfo{journal}{Journal of Geophysical Research}}
  \textbf{\bibinfo{volume}{68}}, \bibinfo{pages}{1233--1263}
  (\bibinfo{year}{1963}).

\bibitem{de-hoffmann_magneto-hydrodynamic_1950}
\bibinfo{author}{De~Hoffmann, F.} \& \bibinfo{author}{Teller, E.}
\newblock \bibinfo{title}{Magneto-hydrodynamic shocks}.
\newblock \emph{\bibinfo{journal}{Phys. Rev.}} \textbf{\bibinfo{volume}{80}},
  \bibinfo{pages}{692--703} (\bibinfo{year}{1950}).

\bibitem{spicer_model_1990}
\bibinfo{author}{Spicer, D.~S.}, \bibinfo{author}{Maran, S.~P.} \&
  \bibinfo{author}{Clark, R.~W.}
\newblock \bibinfo{title}{A model of the pre-sedov expansion phase of supernova
  remnant-ambient plasma coupling and x-ray emission from sn 1987a}.
\newblock \emph{\bibinfo{journal}{The Astrophysical Journal}}
  \textbf{\bibinfo{volume}{356}}, \bibinfo{pages}{549--571}
  (\bibinfo{year}{1990}).

\bibitem{drake_design_2000}
\bibinfo{author}{Drake, R.~P.}
\newblock \bibinfo{title}{The design of laboratory experiments to produce
  collisionless shocks of cosmic relevance}.
\newblock \emph{\bibinfo{journal}{Physics of Plasmas}}
  \textbf{\bibinfo{volume}{7}}, \bibinfo{pages}{4690} (\bibinfo{year}{2000}).

\bibitem{waxer_high-energy_2005}
\bibinfo{author}{Waxer, L.} \emph{et~al.}
\newblock \bibinfo{title}{High-energy petawatt capability for the omega laser}.
\newblock \emph{\bibinfo{journal}{Opt. Photon. News}}
  \textbf{\bibinfo{volume}{16}}, \bibinfo{pages}{30--36}
  (\bibinfo{year}{2005}).

\bibitem{paul_experimental_1965}
\bibinfo{author}{Paul, J. W.~M.}, \bibinfo{author}{Holmes, L.~S.},
  \bibinfo{author}{Parkinson, M.~J.} \& \bibinfo{author}{Sheffield, J.}
\newblock \bibinfo{title}{Experimental observations on the structure of
  collisionless shock waves in a magnetized plasma}.
\newblock \emph{\bibinfo{journal}{Nature}} \textbf{\bibinfo{volume}{208}},
  \bibinfo{pages}{133--135} (\bibinfo{year}{1965}).

\bibitem{fiksel_magnetic_2014}
\bibinfo{author}{Fiksel, G.} \emph{et~al.}
\newblock \bibinfo{title}{Magnetic reconnection between colliding magnetized
  laser-produced plasma plumes}.
\newblock \emph{\bibinfo{journal}{Phys. Rev. Lett.}}
  \textbf{\bibinfo{volume}{113}}, \bibinfo{pages}{105003}
  (\bibinfo{year}{2014}).

\bibitem{niemann_observation_2014}
\bibinfo{author}{Niemann, C.} \emph{et~al.}
\newblock \bibinfo{title}{Observation of collisionless shocks in a large
  current-free laboratory plasma}.
\newblock \emph{\bibinfo{journal}{Geophysical Research Letters}}
  \textbf{\bibinfo{volume}{41}}, \bibinfo{pages}{7413--7418}
  (\bibinfo{year}{2014}).

\bibitem{schaeffer_generation_2012}
\bibinfo{author}{Schaeffer, D.~B.} \emph{et~al.}
\newblock \bibinfo{title}{Generation of magnetized collisionless shocks by a
  novel, laser-driven magnetic piston}.
\newblock \emph{\bibinfo{journal}{Physics of Plasmas}}
  \textbf{\bibinfo{volume}{19}}, \bibinfo{pages}{070702}
  (\bibinfo{year}{2012}).

\bibitem{schaeffer_experimental_2015}
\bibinfo{author}{Schaeffer, D.~B.} \emph{et~al.}
\newblock \bibinfo{title}{Experimental study of subcritical laboratory
  magnetized collisionless shocks using a laser-driven magnetic piston}.
\newblock \emph{\bibinfo{journal}{Physics of Plasmas}}
  \textbf{\bibinfo{volume}{22}}, \bibinfo{pages}{113101}
  (\bibinfo{year}{2015}).

\bibitem{romagnani_observation_2008}
\bibinfo{author}{Romagnani, L.} \emph{et~al.}
\newblock \bibinfo{title}{Observation of collisionless shocks in laser-plasma
  experiments}.
\newblock \emph{\bibinfo{journal}{Physical Review Letters}}
  \textbf{\bibinfo{volume}{101}}, \bibinfo{pages}{025004}
  (\bibinfo{year}{2008}).

\bibitem{kuramitsu_time_2011}
\bibinfo{author}{Kuramitsu, Y.} \emph{et~al.}
\newblock \bibinfo{title}{Time evolution of collisionless shock in
  counterstreaming laser-produced plasmas}.
\newblock \emph{\bibinfo{journal}{Physical Review Letters}}
  \textbf{\bibinfo{volume}{106}}, \bibinfo{pages}{175002}
  (\bibinfo{year}{2011}).

\bibitem{haberberger_collisionless_2011}
\bibinfo{author}{Haberberger, D.} \emph{et~al.}
\newblock \bibinfo{title}{Collisionless shocks in laser-produced plasma
  generate monoenergetic high-energy proton beams}.
\newblock \emph{\bibinfo{journal}{Nat Phys}}  (\bibinfo{year}{2011}).

\bibitem{fox_filamentation_2013}
\bibinfo{author}{Fox, W.} \emph{et~al.}
\newblock \bibinfo{title}{Filamentation instability of counterstreaming
  laser-driven plasmas}.
\newblock \emph{\bibinfo{journal}{Phys. Rev. Lett.}}
  \textbf{\bibinfo{volume}{111}}, \bibinfo{pages}{225002}
  (\bibinfo{year}{2013}).

\bibitem{ross_collisionless_2013}
\bibinfo{author}{Ross, J.~S.} \emph{et~al.}
\newblock \bibinfo{title}{Collisionless coupling of ion and electron
  temperatures in counterstreaming plasma flows}.
\newblock \emph{\bibinfo{journal}{Physical Review Letters}}
  \textbf{\bibinfo{volume}{110}}, \bibinfo{pages}{145005}
  (\bibinfo{year}{2013}).

\bibitem{hewett_physics_2011}
\bibinfo{author}{Hewett, D.~W.}, \bibinfo{author}{Brecht, S.~H.} \&
  \bibinfo{author}{Larson, D.~J.}
\newblock \bibinfo{title}{The physics of ion decoupling in magnetized plasma
  expansions}.
\newblock \emph{\bibinfo{journal}{Journal of Geophysical Research}}
  \textbf{\bibinfo{volume}{116}}, \bibinfo{pages}{12 PP.}
  (\bibinfo{year}{2011}).

\end{thebibliography}


{\footnotesize 
\noindent\textbf{Acknowledgements} We thank the staff of the Omega EP facility for their help in executing these experiments.  Time on the Omega EP faciltiy was funded by the Department of Energy (DOE) through grant DE-NA0002731.  The PIC simulations were conducted on the Hopper supercomputer at the Oak Ridge Leadership Computing Facility at the Oak Ridge National Laboratory, supported by the Office of Science of the DOE under contract number DE-AC05-00OR22725.  This research was also supported by the DOE under contract number DE-SC0008655.
\newline

\noindent\textbf{Author Contributions} W.F., G.F., D.H., A.B., and D.H.B. conceived and designed the experiment. W.F. and G.F. conducted the experiments.  S.X.H. carried out the \textsc{draco} simulations.  W.F. conducted the particle-in-cell simulations using the \textsc{psc} code developed by K.G. and W.F.  The data was analyzed by D.B.S., D.H., and W.F.  The figures were prepared by D.B.S. and the manuscript was written by D.B.S. and W.F.  All authors contributed to the discussion and revision of the manuscript.
\newline

\noindent\textbf{Author Information} Correspondence and requests for materials should be addressed to D.B.S. (dereks@princeton.edu).
}


\end{document}